\documentclass[conference]{IEEEtran}
\usepackage{cite}
\usepackage{amsmath,amssymb,amsfonts}
\usepackage{algorithmic}
\usepackage{graphicx}
\usepackage{textcomp}
\usepackage[table]{xcolor}
\usepackage{xspace}
\usepackage{makecell}
\usepackage{enumitem}
\usepackage{amsmath,amsfonts}
\usepackage{array}
\usepackage{threeparttable}
\usepackage{booktabs}
\usepackage{etoolbox} 

\makeatletter
\long\def\@IEEEsectpunct{.\ \,}
\def\CLASSINPUTinnersidemargin{}
\renewcommand\section{\@startsection{section}{1}{\z@}%
  {-0.5ex plus -0.5ex minus -0.2ex}{0.5ex plus 0.1ex}%
  {\normalfont\normalsize\centering\scshape}}
\renewcommand\subsection{\@startsection{subsection}{2}{\z@}%
  {-0.5ex plus -0.5ex minus -0.2ex}{0.3ex plus 0.1ex}%
  {\normalfont\normalsize\itshape}}
\makeatother

\begin{document}

\title{TeamLLM: Exploring the Capabilities of LLMs for Multimodal Group Interaction Prediction}

\author{\IEEEauthorblockN{Diana Romero, Xin Gao, Daniel Khalkhali, Salma Elmalaki}
\IEEEauthorblockA{\textit{EECS Department} \\
\textit{University of California, Irvine}\\
Irvine, USA \\
dgromer1@uci.edu, xgao10@uci.edu, dkhalkha@uci.edu, salma.elmalaki@uci.edu}
}

\maketitle
\spaceskip=0.25em plus 0.1em minus 0.05em

\begin{abstract}

Predicting group behavior, how individuals coordinate, communicate, and interact during collaborative tasks, is essential for designing systems that can support team performance through real-time prediction and realistic simulation of collaborative scenarios. Large Language Models (LLMs) have shown promise for processing sensor data for human-activity recognition (HAR), yet their capabilities for team dynamics or group-level multimodal sensing remain unexplored. This paper investigates whether LLMs can predict group coordination patterns from multimodal sensor data in collaborative Mixed Reality (MR) environments. We encode hierarchical context---individual behavioral profiles, group structural properties, and temporal activity context---as natural language and evaluate three LLM adaptation paradigms (zero-shot, few-shot, and supervised fine-tuning) against statistical baselines. Our evaluation on 16 groups (64 participants, $\sim$25 hours of sensor data) reveals that LLMs achieve 3.2$\times$ improvement over LSTM baselines for linguistically-grounded behaviors, with fine-tuning reaching 96\% accuracy for conversation prediction while maintaining sub-35ms latency. Beyond performance gains, we characterize the boundaries of text-based LLMs for multimodal sensing conversation prediction succeeds because turn-taking maps to linguistic patterns, while shared or joint attention may require spatial and visual reasoning that text only LLMs cannot capture. We further identify simulation mode brittleness (83\% degradation from cascading context errors) and minimal few-shot sensitivity to example selection strategy. These findings establish guidelines when LLMs are appropriate for CPS/IoT sensing for team dynamics and inform the design of future multimodal foundation models.

\end{abstract}


\section{Introduction}\label{sec:intro}

Group behavior encompasses the emergent patterns of coordination, communication, and interaction among individuals working toward a shared goal. Modeling these dynamics is central to understanding collaborative work, and serves as a prerequisite for enabling both predictive interventions and realistic simulation of collaborative scenarios.
Large Language Models (LLMs) have demonstrated capabilities for sensor data understanding, with recent work showing promise for sensor-in-the-loop reasoning~\cite{ren2025toward} and temporal prediction from event sequences~\cite{shi2023language}. These advances suggest that LLMs can serve as semantic reasoning engines for CPS/IoT applications, interpreting complex sensor patterns and correlation that statistical models struggle to capture. However, existing work primarily targets individual-level sensing tasks~\cite{xu2025exploring}. Whether LLMs can effectively process \textit{multimodal group-level} sensor data, where interactions between individuals create emergent patterns not present in any single stream, remains unexplored.

Group behavior sensing presents unique challenges for foundation models. Collaborative environments such as Mixed Reality (MR) workspaces generate rich multimodal data: gaze vectors revealing attention focus, audio features that capture conversation dynamics, and spatial proximity indicating coordination patterns~\cite{romero2025murmr,chai2025pointpresence}. Predicting how these signals evolve requires reasoning about relational context, capturing not just what each person does but how behaviors interconnect within the group's social dynamics~\cite{pentland2010honest,mcfarland2014network}. Our experiments show that traditional sequence and statistical models struggle to capture the structural coordination patterns that characterize meaningful group behavior, even when provided with rich contextual information.

In this work, we explore the capabilities of LLMs for multimodal group behavior prediction from MR sensor data. We encode hierarchical context (individual behavioral profiles, group network structure, and temporal activity context) as natural language prompts and evaluate three LLM adaptation paradigms spanning the zero-shot to fine-tuned spectrum: (1)~zero-shot prompting tests whether pretrained models possess sufficient social reasoning without task examples, (2)~few-shot in-context learning provides demonstration examples, and (3)~supervised fine-tuning with LoRA adapts Gemma-2B using 402 training samples from 12 collaborative groups.

Our evaluation on 16 groups (64 participants, $\sim$25 hours of sensor data) in an MR group collaborative task reveals that LLMs achieve \textbf{a 3.2$\times$ improvement in predicting group behavior patterns} over statistical baselines when provided with rich context, with fine-tuning reaching 96\% structural similarity for conversation pattern prediction. On the other hand, our results also characterize the boundaries of text-based LLMs for multimodal group sensing: (1)~\textit{modality-dependent success}, where conversation pattern prediction succeeds because turn-taking dynamics map to linguistic patterns encoded during pre-training, while shared attention requires spatial reasoning over 3D coordinates that text tokenization cannot yet fully represent; (2)~\textit{brittleness of the simulation mode}, where context-leveraging models achieve high accuracy with reliable input but degrade 83\% under self-generated context due to cascading error propagation; and (3)~\textit{minimal few-shot sensitivity}, where random example selection performs comparably to sophisticated retrieval strategies (0.5\% difference), suggesting that practitioners can use simple selection without sacrificing performance.

Our contributions include:
\begin{itemize}[topsep=0pt, leftmargin=*, noitemsep]
    \item \textbf{Capability analysis:} We provide the first systematic evaluation of LLM capabilities for group-level multimodal sensor prediction, comparing zero-shot, few-shot, and fine-tuned approaches against statistical baselines.
    \item \textbf{Context encoding:} We demonstrate that encoding hierarchical context (individual profiles, group structure, temporal activity context) as natural language enables LLMs to overcome the context plateau where statistical models fail.
    \item \textbf{Boundary characterization:} We show where text-based LLMs succeed (linguistically-grounded behaviors) versus where they require augmentation (spatial reasoning, autoregressive stability), informing practitioners when LLM-based approaches are appropriate for multi-human CPS/IoT sensing.
\end{itemize}

\section{Related Work}\label{sec:related}

Mobile sensing systems established automatic behavior recognition through integrated accelerometer, acoustic, and location sensing~\cite{miluzzo2008sensing}. Recent work on integrating a sensor-in-the-loop paradigm showcases the capabilities of LLMs for personalized responses from sensor-derived context~\cite{ren2025toward}, but targets individual users rather than collaborative groups. Sociometric badges extended mobile sensing capabilities to group contexts, capturing interaction patterns that correlate with team performance and social dynamics~\cite{kim2012sociometric, zhang2018teamsense}.

Sequential models including RNNs and LSTMs have been widely applied to behavioral prediction~\cite{hochreiter1997long,peng2018aroma}. However, these statistical approaches rely on numerical feature engineering and struggle with semantic context limiting complex behavioral prediction. Thus, recent work explores LLM's for temporal prediction~\cite{jin2024timellm}, causal reasoning of event sequences~\cite{shi2023language}, and powering agents with emergent social behaviors~\cite{park2023generative}. Group behavior prediction introduces unique challenges because of the interweaving of multiple hierarchical levels in the context:  individual traits~\cite{romero2025mocomr}, group structure~\cite{kim2024modeling}, and temporal activity context~\cite{romero2025murmr}.

While LLM-based approaches excel in text-based or simulated environments, their effectiveness for multimodal group behavior in physical and MR settings remains unexplored. Our work addresses this gap by grounding LLM-based behavior modeling in real sensor data from collaborative MR tasks, encoding individual patterns, group structure, and temporal activity context as natural language context.

\section{Methodology}\label{sec:architecture}

\begin{figure*}[t]
  \centering
  \includegraphics[width=\linewidth, trim={0 3.2in 0.3in 1.5in},clip]{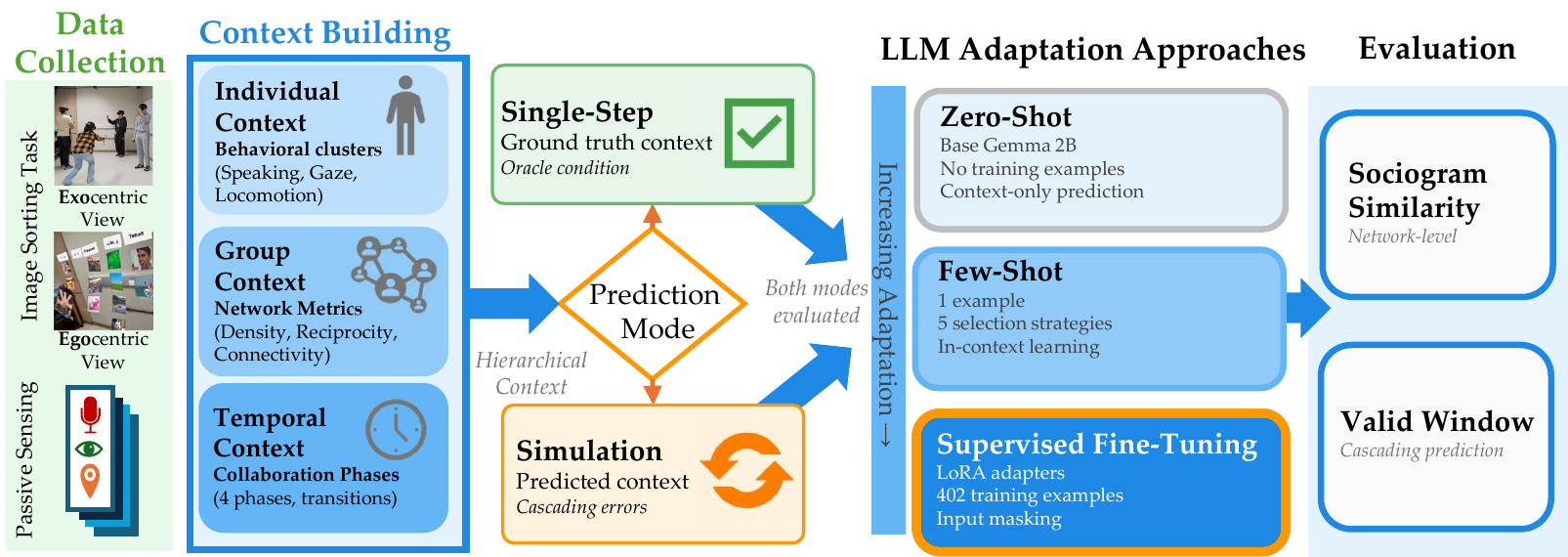}
  \caption{Methodology. Multimodal MR sensor data (gaze, audio, location, task state) are processed into sociograms, then encoded as hierarchical natural language context (individual profiles, group structure, temporal dynamics) for Gemma-2B prediction via zero-shot, few-shot, or LoRA fine-tuning.} 
  \label{fig:framework-overview}
\end{figure*}

Our approach transforms raw multimodal MR sensor data into structured linguistic context for LLM-based group behavior prediction. The pipeline encodes hierarchical context (individual profiles, group structure, temporal dynamics) as natural language, enabling LLMs to reason about structural coordination patterns where statistical models plateau. The complete methodology is shown in Figure~\ref{fig:framework-overview}.

h\subsection{Data and Problem Formulation}\label{sec:data-problem}

\noindent\textbf{Data Collection.} We collected data from 64 participants (age 21-42, mean 24) in 16 groups of 4, using Meta Quest Pro headsets for a collaborative image sorting task~\cite{yang2022towards}. Participants categorized 28 OASIS dataset images into six emotion categories~\cite{kurdi2017introducing}, encouraging decision-making, communication, and social coordination.  The IRB-approved study captured four synchronized sensor channels into 32-second windows, a duration empirically validated through ablation studies showing behavioral prediction accuracy plateaus beyond 30 seconds~\cite{romero2024groupbeamr} and consistently adopted across prior analyses of this collaborative MR dataset~\cite{chandio2026sensors, romero2025mocomr, romero2025murmr}:
\begin{enumerate}[noitemsep,topsep=0pt,leftmargin=*]
  \item \textbf{Gaze}: Eye tracking vectors at 72Hz, processed into shared attention metrics via joint attention detection~\cite{moore2014joint};
  \item \textbf{Audio}: Microphone streams for speaker diarization and turn-taking analysis;
  \item \textbf{Position}: 6-DoF headset tracking for relative interpersonal distances;
  \item \textbf{Task State}: Application logs capturing image selections and category assignments.
\end{enumerate}

\noindent\textbf{Sociogram Representation.} Instead of predicting low-level sensor values, we predict structural group coordination represented as time-evolving sociograms $G_t$, where nodes are individuals and edges represent interaction strength.

Specifically, for each time window ($T=32$ seconds), we construct three concurrent sociograms for each group $G$ in window $t$:

\begin{enumerate}[topsep=0pt, leftmargin=*, noitemsep]
    \item \textbf{Conversation Sociogram $G^\text{conv}$:} is a directed graph where the edge weight $w_{i,j}$ represents the normalized amount of speaking time person $i$ directed toward person $j$.
    
    \item \textbf{Proximity Sociogram $G^\text{prox}$:} is an undirected graph where the edge weight $w_{i,j}$ represents the total duration person $i$ spent in close-proximity threshold (e.g., 1.5 ft) of person j.
    
    \item \textbf{Shared Attention Sociogram $G^\text{att}$:} is an undirected graph where the edge weight $w_{i,j}$ represents the duration with which person~$i$ shared attention on the same object or area as person~$j$. 
\end{enumerate}

This structural perspective is critical: models can achieve high element-wise accuracy while missing coordination patterns like turn-taking shifts or subgroup formations, as discussed in~\S\ref{sec:persistence}.

\subsection{Context Encoding and LLM Pipeline}\label{sec:encoding-pipeline}

\textbf{Multi-Level Context Encoding.} We serialize multimodal, network-level context into natural language prompts, enabling LLMs to leverage semantic reasoning. The LLM predicts $G_{t+1}$ based on three context levels from window $t$: 
\begin{enumerate}[noitemsep,topsep=0pt,leftmargin=*]
    \item \textbf{Individual Behavioral Profiles} ($C_t^{indiv}$): stable traits encoded via MoCoMR clustering~\cite{romero2025mocomr} using unsupervised GMM analysis of speaking frequency, gaze patterns, and locomotion dynamics, with natural language descriptors (e.g., ``frequent talker,'' ``high gaze activity'')
    \item \textbf{Group Structural Properties} ($C_t^{group}$): network-level state including density, reciprocity, eigenvector centrality, and clustering coefficients~\cite{mcfarland2014network}, combined via PCA-weighted fusion~\cite{romero2025mocomr}
    \item \textbf{Temporal Dynamics} ($C_t^{temp}$): phase clustering via MURMR~\cite{romero2025murmr} (e.g., exploration, active discussion, consensus), phase-level metrics, and 5-window interaction history.
\end{enumerate}

Context serialization is illustrated in Table~\ref{tab:context-example}.

\begin{table}[t]

\caption{Multi-level context example. Individual clusters remain constant; group and temporal contexts update every 32 seconds.}
\label{tab:context-example}
\begin{tabular}{@{}p{1.2cm}p{6.5cm}@{}}
\toprule
\textbf{Level} & \textbf{Representation} \\ \midrule
Individual & Speaking: Frequent Talker (Cluster 0, 119/session) \\
           & Gaze: High Activity (Cluster 2, diverse focus) \\
           & Locomotion: Dynamic (Cluster 2, high speed) \\ \midrule
Group & Conversation: $\rho=0.35$, reciprocity=0.42 \\
      & Proximity: $\rho=0.28$, clustering=0.31 \\
      & Shared Attention: $\rho=0.15$, clustering=0.09 \\ \midrule
Temporal & Phase: Animated Collaboration (Cluster 1) \\
         & Duration: 5 consecutive windows \\
         & Trends: Density $\uparrow$ (0.15$\rightarrow$0.22), Reciprocity $\uparrow$ \\ \bottomrule
\end{tabular}
\end{table}

\noindent\textbf{Theoretical Framing.} This hierarchical encoding aligns with the \textit{Input-Process-Output (IPO)} model from team science\cite{ilgen2005teams} and the \textit{Shared Mental Model (SMM)} framework \cite{mathieu2000influence}: individual profiles capture input factors (member characteristics), group structure reflects emergent processes (interaction patterns), and temporal dynamics track output trajectories. By encoding these levels as natural language, the LLM can leverage pre-trained knowledge about group dynamics for prediction. The explicit mapping between team science constructs and our system components is shown in Table~\ref{tab:ipo-mapping}.

\begin{table}[t]
\centering
\caption{Mapping team science constructs to system components.}
\label{tab:ipo-mapping}
\begin{tabular}{@{}p{1.4cm}p{2.3cm}p{3.8cm}@{}}
\toprule
\textbf{IPO} \newline \textbf{Component} & \textbf{Description} & \textbf{Technical} \newline \textbf{Operationalization} \\
\midrule
Input & Static team composition & Individual profiles ($C_t^{indiv}$) \\
Process & Team interactions and coordination & Sociograms ($G_t$) and context encoding ($C_t^{temp}$, $C_t^{group}$) \\
Output & Emergent structural states & Predicted sociogram ($G_{t+1}$) \\
\bottomrule
\end{tabular}

\end{table}

\begin{figure}[t]
  \centering
  \includegraphics[width=\linewidth, trim={0 1.6in 8.2in 1.5in},clip]{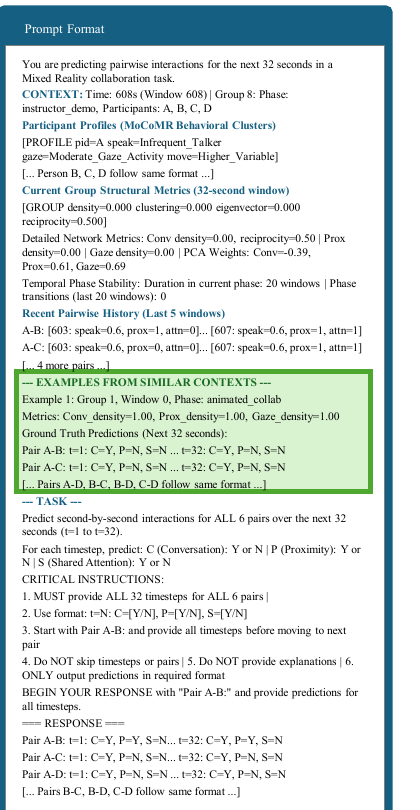}
  \caption{Prompt structure showing participant profiles, group metrics, temporal phase, pairwise history, and few-shot examples (green, omitted in zero-shot).}
  \label{fig:prompt}

\end{figure}

\noindent\textbf{Prompt Structure.} The prompt $P_t$ comprises seven components:

$$P_t = [I \mid C_t^{\text{temp}} \mid C_t^{\text{indiv}} \mid C_t^{\text{group}} \mid H_t^{\text{pair}} \mid E_t^{\text{event}} \mid \text{FMT}]$$

\noindent comprising: (1)~task instructions $I$, (2)~temporal context $C_t^{\text{temp}}$ (phase, stability, trends), (3)~individual profiles $C_t^{\text{indiv}}$ (MoCoMR clusters~\cite{romero2025mocomr}), (4)~group structural metrics $C_t^{\text{group}}$ (density, reciprocity, centrality, clustering), (5)~pairwise interaction history $H_t^{\text{pair}}$ (last 5 windows), (6)~event timeline $E_t^{\text{event}}$ (last 10 windows), and (7)~output format $\text{FMT}$ as illustrated in Figure~\ref{fig:prompt}.

The LLM generates timestep-by-timestep binary predictions in structured text format. For each participant pair and each second $s \in [t+1, t+32]$, the model predicts three binary interaction indicators using the notation \texttt{t=s: C=[Y/N], P=[Y/N], S=[Y/N]}, where \texttt{C}, \texttt{P}, and \texttt{S} denote conversation, proximity, and shared attention, respectively. A parsing system (\texttt{ResponseParser}) extracts these predictions using pattern matching with multiple fallback strategies to handle output variability, converting text responses into binary adjacency matrices $G_{t+1} = \{G_{t+1}^{\text{conv}}, G_{t+1}^{\text{prox}}, G_{t+1}^{\text{attn}}\}$ for quantitative evaluation.

\noindent\textbf{Inference Modes.} We evaluate two deployment modes:  \textit{Single-Step Mode}: single-step prediction $G_t \rightarrow G_{t+1}$ with ground truth context, isolating the model's reasoning capacity from input errors; and \textit{Simulation Mode}: autoregressive prediction $G_t \rightarrow G_{t+1} \rightarrow G_{t+2} \rightarrow \cdots$ where predicted sociograms feed back as context, testing robustness to compounding errors.

\noindent\textbf{LLM Adaptation Paradigms.} We evaluate three configurations: (1)~\textit{Zero-Shot}: Gemma-2B-IT receives structured context ($\sim$4,200 tokens) without task-specific examples, testing whether pre-trained knowledge suffices for group behavior reasoning; (2)~\textit{Few-Shot} ($k=1$): single demonstration example ($\sim$4,552 additional tokens) due to Gemma-2B's 8,192-token limit, comparing random, phase-similar, and diversity-based example selection strategies; (3)~\textit{Supervised Fine-Tuning}: LoRA adaptation~\cite{hu2022lora} with $r=16$, $\alpha=32$, targeting attention projections (2.5M trainable parameters, 0.10\% of 2.6B), trained on Groups 1-12 for 3 epochs with AdamW ($lr=10^{-4}$), computing loss only on prediction tokens.

\subsection{Evaluation Protocol}\label{sec:eval-protocol}

\noindent\textbf{Dataset and Experimental Design.} We use data from 16 groups (\S\ref{sec:data-problem}): sessions of 4.8--18.4 minutes (mean: 9.7) segmented into 32-second windows with 16-second stride~\cite{romero2024groupbeamr, chandio2026sensors, romero2025mocomr, romero2025murmr}, yielding 447 active windows ($\sim$25~hours total). Prompting approaches (zero-shot, few-shot) are evaluated on all 16 groups; few-shot examples selected from other groups using random, few-shot examples ($k=1$) are selected from other groups via three strategies: random sampling, cosine similarity in temporal phase embeddings to retrieve contextually matched demonstrations, or k-means++ initialization to maximize interaction pattern coverage. Fine-tuning is trained on Groups 1-12, evaluated on held-out Groups 13-16. Ablation studies use a single group to isolate context component contributions. All paradigms are evaluated under both single-step and simulation modes (evaluating six configurations in total).

\noindent\textbf{Baselines.} We compare against simple baselines: (1)~\textit{Persistence}: repeats last sociogram; (2)~\textit{Temporal Smoothing}: averages previous $N \in \{3,5\}$ sociograms; (3)~\textit{Stratified Random}: samples edges matching empirical interaction rates. We also compare against a statistical baseline: Bidirectional LSTM (2048 units) with attention over the same multi-level context, using continuous features instead of natural language, trained on Groups 1-12.

\noindent\textbf{Metrics.}\label{sec:eval-metrics} Our primary metric is \textit{Sociogram Similarity}: weighted Jaccard similarity between predicted and ground truth sociograms, capturing structural coordination patterns~\cite{mcfarland2014network, pentland2010honest}. Persistence achieves 95\% pairwise accuracy but only 6\% sociogram similarity (\S\ref{sec:persistence}), motivating structural metrics. Secondary metrics include: (1)~\textit{Valid Window Rate}: proportion achieving $\geq$80\% accuracy; (2)~\textit{Network Property Preservation}: Pearson correlation between predicted and ground truth network metrics (density, reciprocity, clustering), measuring whether models capture relative structural trends across windows independent of absolute scale.

\section{Experimental Evaluation}\label{sec:experiments}

We evaluate LLM-based group behavior prediction against statistical baselines, demonstrating that statistical models plateau at 29\% sociogram similarity regardless of context complexity, while LLMs achieve 96\% for conversation prediction in single-step mode.

\subsection{Limitations of Statistical Models}\label{sec:stat-limits}

We evaluate statistical approaches to establish baseline capabilities and motivate our LLM-based approach.

\textbf{The Persistence Problem.}\label{sec:persistence} Simple persistence (repeating the last sociogram) achieves deceptively high element-wise accuracy by exploiting temporal autocorrelation, but fails to capture coordination dynamics (Table~\ref{tab:persistence_problem}). Persistence achieves F1 $>$ 0.89 and MCC of 0.891, yet sociogram similarity remains at 0.062. This gap arises because persistence correctly predicts the majority class due to high autocorrelation ($\rho \approx 0.53$-0.73), but misses turn-taking shifts, attention reallocations, and spatial regrouping. Conversation exhibits high F1 (0.984) but near-zero MCC ($-$0.008) due to extreme class imbalance (99.76\% active pairs). \textbf{This motivates sociogram similarity as our primary metric}, as only structural metrics capture coordination dynamics versus class imbalance exploitation.

\begin{table}[t]

\centering
\caption{Persistence baseline achieves strong element-wise metrics but poor sociogram similarity, revealing failure to capture coordination changes.}
\label{tab:persistence_problem}
\begin{tabular}{lcccc}
\toprule
\textbf{Metric} & \textbf{Conv.} & \textbf{Prox.} & \textbf{Att.} & \textbf{Overall} \\
\midrule
F1 Score & 0.984 & 0.896 & 0.936 & 0.955 \\
MCC & $-$0.008 & 0.840 & 0.887 & 0.891 \\
\midrule
\rowcolor[gray]{0.9}
\multicolumn{5}{c}{\textit{Sociogram Similarity (Weighted Jaccard)}} \\
Single-Step Mode & 0.068 & 0.045 & 0.074 & 0.062 \\
\bottomrule
\end{tabular}

\end{table}

\textbf{The Context Plateau.}\label{sec:context-plateau} We trained LSTM models (2048 units, 30 epochs) across five context configurations: individual features, group concatenation, group + sociogram history, group + behavioral clusters, and full context. Results reveal that \textbf{performance plateaus at $\sim$29\% weighted Jaccard regardless of context complexity}, with $<$2\% variation across configurations (Table~\ref{tab:context_plateau})
This plateau occurs despite 55-56\% training accuracy on sensor-value predictions, revealing an accuracy-similarity gap: \textbf{models optimize for individual prediction without preserving group coordination structure.}

\begin{table}[t]

\centering
\caption{LSTM plateaus at $\sim$29\% weighted Jaccard regardless of context complexity ($<$2\% variation), demonstrating architectural rather than data limitations.}
\label{tab:context_plateau}
\begin{tabular}{lcc}
\toprule
\textbf{Context Configuration} & \textbf{Overall WJ} & \textbf{$\Delta$} \\
\midrule
Individual Only & 29.3\% & --- \\
Group + Sociogram History & 28.2\% & -1.1\% \\
Full Context & 28.7\% & -0.6\% \\
\bottomrule
\end{tabular}

\end{table}

These findings suggest that the limitation is architectural: LSTMs lack semantic reasoning to interpret coordination patterns like turn-taking and role-based behaviors. This motivates LLM-based approaches that leverage natural language context representations.


\subsection{LLM Performance}\label{sec:llm-performance}

We evaluate whether LLMs can break through the 29\% statistical ceiling, as reported in \S\ref{sec:stat-limits}, using three configurations: zero-shot, few-shot (k=1), and LoRA fine-tuning.

\textbf{Single-Step Mode Results.} Sociogram similarity in single-step mode (ground truth context) is presented in Table~\ref{tab:main_sociogram_results}. SFT achieves 0.958 weighted Jaccard for conversation—a \textbf{3.2x improvement} over LSTM (0.298). Few-shot with temporally matched examples (selected by cosine similarity in MURMR's temporal phase embeddings~\cite{romero2025murmr}) achieves 0.582 average similarity without task-specific training, outperforming LSTM (0.231). Zero-shot underperforms all baselines (0.101), confirming that pre-trained knowledge alone is insufficient. SFT exhibits modality-dependent specialization: near-optimal conversation (0.958) but weak proximity (0.104), suggesting semantic patterns are easier to capture than spatial interactions.

\begin{table}[t]
  
  \centering
  \caption{Single-Step mode sociogram similarity. LLMs achieve 3.2x improvement over LSTM. Shared attention (--) suggests visual grounding beyond text-based LLM capabilities is required.}
  \label{tab:main_sociogram_results}
  \begin{tabular}{lcccc}
  \toprule
  \textbf{Method} & \textbf{Conv} & \textbf{Prox} & \textbf{Avg} & \textbf{vs LSTM} \\
  \midrule
  \rowcolor[gray]{0.9}
  \multicolumn{5}{c}{\textit{Baselines}} \\
  Persistence & 0.068 & 0.045 & 0.062 & --- \\
  LSTM + Full Context & 0.298 & 0.215 & 0.231 & 1.0x \\
  \midrule
  \rowcolor[gray]{0.9}
  \multicolumn{5}{c}{\textit{LLM Approaches}} \\
  Zero-Shot & 0.067 & 0.188 & 0.101 & 0.4x \\
  Few-Shot (Similar, k=1) & 0.785 & 0.474 & 0.582 & 2.5x \\
  SFT (LoRA) & \textbf{0.958} & 0.104 & 0.395 & \textbf{3.2x} \\
  \bottomrule
  \end{tabular}
\end{table}

\textbf{Computational Efficiency.} We evaluate inference on an NVIDIA RTX 4060 Ti (consumer hardware). Despite 7.6x context increase (237 to 1,799 characters), TTFT remains stable at $<$35ms and total inference time \textit{decreases} by 19.7\% (4.15s to 3.33s). Memory increases by only 0.7\% (5.02GB to 5.05GB). Scalability tests across 14x context range (237-3,328 chars) show sub-linear scaling with TTFT stable at 32-34ms, confirming context-rich prompts deploy efficiently on consumer hardware.

\textbf{Simulation Mode: Error Propagation.} Simulation mode reveals brittleness to cascading errors (Table~\ref{tab:simulation_results}). Few-shot conversation similarity collapses from 0.785 to 0.007 (99.1\% degradation); SFT degrades from 0.958 to 0.165 (82.8\%). Initial mispredictions corrupt group context metrics (density, reciprocity), biasing subsequent predictions. Conversation predictions fail by cascade depth 2 (half-life $\leq$2 windows), indicating binary failure rather than gradual decay. Proximity remains stable (0.425), suggesting spatial patterns are less context-dependent. The stratified random baseline outperforms LLMs in simulation (0.993 vs 0.007), demonstrating that context-free models avoid error accumulation but miss temporal dynamics. This reveals a practical tradeoff: models that rely on rich behavioral context achieve strong single-step predictions but are sensitive to errors in that context, while simpler models that ignore context remain stable but cannot capture how group behavior evolves over time.

\begin{table}[t]

\centering
\caption{Simulation mode performance. LLM approaches exhibit severe degradation (83-99\%) due to cascading error propagation.}
\label{tab:simulation_results}
\begin{tabular}{lccc}
\toprule
\textbf{Method} & \textbf{Conv} & \textbf{Prox} & \textbf{Degradation} \\
\midrule
Few-Shot (Similar) & 0.007 & 0.425 & -99.1\% \\
SFT & 0.165 & 0.104 & -82.8\% \\
Stratified Random & 0.993 & 0.145 & --- \\
\bottomrule
\end{tabular}
\end{table}

\subsection{Analysis}\label{sec:analysis}

\textbf{Modality-Dependent Boundaries.} Ablation across context levels reveals where text-based LLMs succeed and fail: (1)~\textit{Conversation} achieves 97-100\% similarity regardless of context level, demonstrating that turn-taking dynamics map naturally to linguistic patterns encoded during LLM pre-training; (2)~\textit{Proximity} improves from 0\% to 6\% with full context (t=3.42, p=0.001, Cohen's d=0.63), benefiting from behavioral descriptions but limited by spatial reasoning requirements; (3)~\textit{Shared attention} remains at 0\% across all configurations, including with full context, class rebalancing, and fine-tuning. This persistent failure indicates limitations of text only LLMs as joint gaze prediction requires reasoning over 3D coordinates and visual object semantics that discrete text tokenization cannot represent. These results provide guidelines on when text-based LLMs are appropriate (linguistically-grounded behaviors) versus when vision-language models are required (geometric reasoning). These findings motivate for the exploration of a tiered sensing strategies: audio-only for conversation, audio + position for proximity, and gaze with semantic object recognition for shared attention.

\textbf{Few-Shot Example Selection.} Due to Gemma-2B's 8K context limit and per-example costs (4,552 tokens), we evaluated single-example strategies (Table~\ref{tab:few_shot_strategies}). Phase-similar selection achieves only 0.5\% improvement over random (0.582 vs 0.579), suggesting Gemma-2B extracts reasoning patterns from any appropriate example. Given negligible benefit and computational overhead ($O(N \log N)$ vs $O(1)$), random selection offers the best cost-benefit tradeoff.

\begin{table}[t]

\centering
\caption{Few-shot strategy comparison (k=1). Phase-similar selection provides minimal benefit over random sampling.}
\label{tab:few_shot_strategies}
\begin{tabular}{lcc}
\toprule
\textbf{Strategy} & \textbf{Similarity} & \textbf{Complexity} \\
\midrule
Random & 0.579 & $O(1)$ \\
Phase Similar & 0.582 & $O(N \log N)$ \\
Diverse & 0.554 & $O(N^2)$ \\
\bottomrule
\end{tabular}

\end{table}

\textbf{Fine-Tuning Configuration.} LoRA adaptation ($r=16$, $\alpha=32$) yields 2.5M trainable parameters (0.10\% of base model), trained for 3 epochs on Groups 1-12, achieving 99.04\% token accuracy. A critical issue was sequence length: default 1,024-token limit caused truncation of our 4,552-token examples, producing incomplete outputs (2/6 pairs). Setting \texttt{model\_max\_length=8192} resolved this, achieving 6/6 complete pairs. We apply loss masking on prediction tokens only, forcing the model to learn context-to-prediction mapping rather than input reconstruction.

\section{Conclusion and Future Work}\label{sec:conclusion}

We demonstrated that LLMs can leverage hierarchical contextual information to predict group coordination patterns in collaborative MR environments, achieving 3.2$\times$ improvement over statistical baselines (96\% vs 29\% sociogram similarity for conversation). The context plateau, where LSTMs achieve constant performance regardless of context richness, suggests an architectural limitation: statistical models process behavioral clusters, network metrics, and temporal phases as numerical features without understanding their semantic relationships. LLMs are able to overcome this through compositional reasoning about how social roles interact with group states.

\textbf{Characterizing LLM Boundaries.} Our evaluation reveals where text-based LLMs succeed and fail for multimodal group sensing. Conversation prediction achieves 96\% accuracy because turn-taking and speaker dynamics map naturally to linguistic patterns that LLMs encode during pre-training. Shared attention prediction fails completely (0\% recall) because joint gaze requires spatial reasoning over 3D coordinates and visual object semantics that discrete text tokenization cannot represent. This insight is further supported by the performance of shared attention prediction which remains at 0\% regardless of context richness, class rebalancing, or fine-tuning. These findings provide guidelines on when text-based LLMs are appropriate for CPS/IoT sensing (linguistically-grounded behaviors) versus when vision-language or spatial foundation models are required (geometric and visual reasoning).

\textbf{Future Work.} Vision-language models with coordinate embeddings could address spatial reasoning limitations. Hybrid architectures combining LLM semantic reasoning with statistical error buffering, or constrained decoding with periodic ground truth injection, may extend simulation horizons. Results motivate tiered sensing: audio-only for conversation ($>$95\% accuracy), audio + position for proximity, and gaze with semantic object recognition for shared attention. The sub-35ms inference latency confirms feasibility for real-time coordination support in collaborative MR systems.

\section*{Acknowledgment}

This work is supported by the U.S. National Science Foundation (NSF) under grant number 2339266.

\begingroup
\renewcommand{\footnotesize}{\fontsize{8.5}{8}\selectfont}
\bibliographystyle{IEEEtran}
\bibliography{paper}
\endgroup

\end{document}